\documentclass[prd,superscriptaddress,amsfonts,amssymb,amsmath,showpacs,twocolumn]{revtex4-2}
\usepackage{amsfonts,amsmath,units,graphicx,verbatim,color,subfig,graphicx,bm,mathrsfs,lipsum,hyperref,cleveref}
\usepackage{booktabs}
\usepackage[normalem]{ulem}  
\definecolor{dgreen}{RGB}{0, 204, 0}
\usepackage{multirow}

\usepackage{bm}
\usepackage{amsfonts}
\usepackage{latexsym}
\usepackage{textcomp}
\usepackage{orcidlink}

\hypersetup{colorlinks,citecolor=red}

\newcommand{\be}{\begin{equation}}
\newcommand{\ee}{\end{equation}}
\newcommand{\bea}{\begin{eqnarray}}
\newcommand{\eea}{\end{eqnarray}}

\DeclareUnicodeCharacter{2212}{-}
\usepackage[none]{hyphenat}

\usepackage{cellspace}
\begin{document}

\title{Bayesian Inference of Neutron Star Properties in $f(Q)$ Gravity Using NICER Observations}

\author{Sneha Pradhan\orcidlink{0000-0002-3223-4085}}
\email{snehapradhan2211@gmail.com}
\affiliation{Department of Mathematics, Birla Institute of Technology and
Science-Pilani,\\ Hyderabad Campus, Hyderabad-500078, India}

\author{N. K. Patra\orcidlink{0000-0003-0103-5590}} \email{nareshkumarpatra3@gmail.com} \affiliation{School of Science and Engineering, The Chinese University of Hong Kong, Shenzhen (CUHK-Shenzhen), Guangdong 518172, China}

\author{Kai Zhou\orcidlink{0000-0001-9859-1758}}
\email{zhoukai@cuhk.edu.cn}
\affiliation{School of Science and Engineering,
The Chinese University of Hong Kong, Shenzhen (CUHK-Shenzhen),
Guangdong 518172, China}
\affiliation{School of Artificial Intelligence, The Chinese University of Hong Kong, Shenzhen (CUHKShenzhen), Guangdong, 518172, China}

\author{P.K. Sahoo\orcidlink{0000-0003-2130-8832}}
\email{pksahoo@hyderabad.bits-pilani.ac.in}
\affiliation{Department of Mathematics, Birla Institute of Technology and
Science-Pilani,\\ Hyderabad Campus, Hyderabad-500078, India}

\date{\today}

\begin{abstract}

In this work, we investigate neutron stars (NSs) in the strong field regime within the framework of symmetric teleparallel $f(Q)$ gravity, considering three representative models: linear, logarithmic, and exponential. While Bayesian studies of NS observations are well established in general relativity and curvature based modified gravity theories, such analyses in $f(Q)$ gravity remain largely unexplored. For the first time we perform a Bayesian inference analysis by confronting theoretical NS mass-radius predictions with NICER observations of PSR J0030+0451, PSR J0740+6620, PSR J0437+4715, and
PSR J0614+3329 in the background of nonmetricity based gravity. The dense matter equation of state is fixed to DDME2 in order to isolate the effects of modified gravity on NS structure. Our results show that the exponential $f(Q)$ model is statistically preferred over the linear and logarithmic cases, as confirmed by Bayes factor comparisons, and exhibits well-constrained. For this model, we obtain a radius and tidal deformability at $1.4\,M_\odot$ of $R_{1.4} = 11.27^{+0.53}_{-0.36}\,\mathrm{km}$ and $\Lambda_{1.4} = 156.95^{+84.02}_{-41.73}$, respectively, consistent with current observational constraints. Remarkably, all three constrained models predict maximum neutron star masses reaching $M_{\max} \simeq 2.98\,M_{\odot}$, with the $95\%$ confidence regions extending into the lower mass gap ($\sim 2.5$--$5\,M_{\odot}$). This mass-gap prediction emerges naturally from the Bayesian-constrained parameter space. 
These results highlight the potential of
NSs as powerful probes of symmetric teleparallel gravity in the strong field regime.

\end{abstract}

\maketitle
\section{Introduction}\label{sec:1}
Einstein’s General Theory of Relativity (GR) \citep{Einstein:1916} provides a remarkably successful theory of gravitation, accurately explaining gravitational phenomena from Solar System tests to cosmological scales. In this framework, gravity is not treated as a fundamental force, but rather emerges from the curvature of spacetime induced by the presence of matter and energy \citep{Wald:1984rg}. Despite its empirical success, GR encounters conceptual and observational challenges when applied to the entire Universe \citep{Will_2014}. In particular, the appearance of spacetime singularities signals a breakdown of the classical theory, while the phenomena attributed to dark matter and dark energy require the introduction of unknown components within the standard GR framework. These issues strongly motivate the exploration of modified theories of gravity (MG) \citep{2012PhR...513....1C}.

In recent years, there has been growing interest in MG as possible explanations for the observed late-time acceleration of the Universe. This accelerated expansion has been firmly established by several independent observational probes, including Type Ia supernovae, measurements of cosmic microwave background anisotropies, and large-scale structure surveys \citep{Perlmutter:1999,Riess:1998,Spergel:2007,Tegmark:2005}. Although the cosmological constant offers the simplest description consistent with these observations, its interpretation within the framework of quantum field theory remains unsatisfactory. In particular, the observed value of the cosmological constant is much smaller than the theoretical predictions, giving rise to the well-known fine-tuning and coincidence problems \citep{Sahni:2000,Padmanabhan:2003}. These unresolved issues have motivated a wide range of alternative ideas, including dynamical dark energy scenarios and modifications of gravity itself \citep{Linder:2008,Frieman:2008,Caldwell:2009}.

In the standard Riemannian formulation of GR, gravity is described as a manifestation of the curvature of the spacetime manifold. From an effective field theory perspective, it is natural to expect that higher-order geometric contributions may become relevant at high energy scales or in strong-gravity regimes. This expectation has motivated several extensions of GR, among which two widely studied approaches are the introduction of additional scalar degrees of freedom and the generalization of the Einstein–Hilbert action by replacing the Ricci scalar 
$R$ with a nonlinear function $f(R)$ \citep{DeFelice:2010,Sotiriou:2010,Capozziello:2011,Nojiri:2011}. Alternatively, GR admits equivalent formulations based on different choices of affine connections, offering distinct geometric interpretations of gravitation while preserving dynamical equivalence at the level of field equations \citep{PhysRevD.101.024053,Aldrovandi:2013}. One such formulation is the teleparallel equivalent of General Relativity (TEGR), where gravity is encoded in spacetime torsion rather than curvature, the tetrad field serves as the fundamental dynamical variable, and the gravitational action is constructed from the torsion scalar 
$T$, with the curvature identically vanishing \citep{Aldrovandi:2013,Maluf:2013}. A further geometric reformulation is provided by the symmetric teleparallel equivalent of General Relativity (STEGR), in which both curvature and torsion vanish, and gravity is entirely described by the nonmetricity of spacetime through the disformation tensor \citep{PhysRevD.98.044048,Zhao:2021zab}. In this framework, the gravitational interaction is governed by the nonmetricity scalar 
$Q$, which forms the basis for the construction of modified theories such as $f(Q)$ gravity. These theories offer a novel and geometrically distinct extension of GR, with promising applications in cosmology and astrophysics.
In this work, we investigate the internal structure of neutron stars (NSs) within the framework of $f(Q)$ gravity by considering three representative functional forms of the theory, namely linear, logarithmic and exponential models \citep{wang/2022,log,gaurav}. These choices encompass both linear and nonlinear extensions of the gravitational action with respect to the nonmetricity scalar $Q$, leading to modified field equations that govern stellar equilibrium. By systematically analyzing these models, we aim to assess how departures from general relativity (GR) influence the macroscopic properties of NSs, such as their mass–radius relations and internal profiles. MG theories such as the $f(Q)$ models considered here are well motivated, as they can potentially explain gravitational phenomena without invoking additional dark matter. However, any viable theory must satisfy broad astrophysical and observational constraints. Notably, $f(Q)$ models have been extensively studied in cosmology and shown to fit data well, reproducing late-time cosmic acceleration, dynamical dark energy behavior, and stable phase-space evolution \citep{simu,log1}. Crucially, a complete theory must also address the strong-field regime. NSs are the compact objects with intense gravitational fields and extreme physics provide a powerful test of $f(Q)$ gravity beyond cosmology.

In parallel with advances in gravitational theory, rapid progress in neutron-star observations has established compact objects as powerful probes of strong-field gravity. The dense matter equation of state (EoS) governs neutron-star properties such as the mass–radius relation, tidal deformability, oscillation modes, and internal structure, which in turn determine observable signatures in gravitational-wave (GW), electromagnetic, and thermal emissions \citep{Hinderer:2008,Hinderer:2010,Andersson:1998,Sotani:2007,Lattimer:2001,Ozel:2016}. Over the past decade, precise measurements have significantly constrained the EoS. In particular, the discovery of NSs with masses near $2M_{\odot}$ has ruled out many overly soft EoS models, imposing stringent limits on matter at supranuclear densities \citep{Demorest:2010,Antoniadis:2013,Cromartie:2019,Fonseca:2021}. Further constraints have been provided by observations with the NS Interior Composition Explorer (NICER), which has provided the first simultaneous mass–radius measurements of rotation-powered millisecond pulsars through relativistic X-ray pulse-profile modeling. In particular, analyses of PSR J0030+0451 and PSR J0740+6620 favor radii in the range $R \sim 12$–$14\mathrm{km}$ for NSs with masses near $1.4M_{\odot}$ and $\sim 2M_{\odot}$, respectively, thereby ruling out both extremely soft and overly stiff EoS \citep{Gendreau:2017,Miller:2021,Riley:2021,Karan_2025}. Complementary constraints arise from GW observations of binary neutron-star mergers, most notably GW170817, which impose stringent upper bounds on the dimensionless tidal deformability of a canonical $1.4M_{\odot}$ NS, $\Lambda_{1.4} \lesssim 800$, further narrowing the space of viable EoS models \citep{Abbott:2018}.

Within GR, a number of studies have combined multimessenger NS observations to place stringent constraints on stellar structure and dense-matter physics. More recently, similar strategies have been applied to MG, where NS observables such as maximum mass–radius relations and tidal deformabilities, as sensitive probes of deviations from GR and of additional model parameters \citep{Berti:2015,CapozzielloNS:2016,Staykov:2014,Moreno_2025}. A considerable number of studies have investigated the dynamical stability of neutron stars within various modified theories of gravity, including 
$f(R)$ gravity \citep{Cheoun_2013, Astashenok_2013,Moreno_2025}, Rainbow–Rastall gravity \citep{Hendi_2016}, and 
$f(T)$ gravity \citep{Vilhena_2023}. These works have provided valuable theoretical insights into how deviations from general relativity can influence neutron-star stability. However, in most of these analyses, the model parameters are chosen phenomenologically often through trial-and-error procedures, without being systematically constrained by observational data. Consequently, the astrophysical viability of such models remains largely untested against current observations. This motivates the need for a systematic investigation in which modified-gravity neutron-star models are tested against observational constraints, thereby bridging the gap between theoretical consistency and astrophysical viability. In this context, Bayesian inference has emerged as a powerful framework for confronting theoretical models with observational data. Bayesian analyses combining NICER mass–radius measurements with GW constraints from binary NS mergers have yielded statistically robust constraints on the dense-matter EoS within GR \citep{Raaijmakers:2021,Essick:2020,Jiang_2023,PhysRevLett.121.062701,article}. These methods have subsequently been extended to MG, notably $f(R)$ and related theories, enabling the joint inference of the EoS and gravity-sector parameters by maintaining consistency with NICER observations, GW signals, and massive pulsars \citep{2023PhRvD.107l4045N,PhysRevD.109.064048}. Such studies demonstrate that compact-object observations can impose stringent bounds on departures from GR.
Despite this progress, most observationally driven constraints have focused on curvature-based extensions of gravity, STEGR has received comparatively little attention from an observational standpoint. In particular, although $f(Q)$ gravity possesses the distinct geometric formulation, where gravitational dynamics are encoded in spacetime influenced by non-Riemannian geometry (non-metricity), a systematic Bayesian inference of its model parameters using NS observables remains largely unexplored. This motivates the present work, in which NSs data are employed to directly constrain the parameter space of 
$f(Q)$ gravity beyond cosmological tests.
In this work, we address this gap by performing a comparative analysis of NSs in the strong-gravity regime within the framework of  $f(Q)$ gravity, considering three representative models - linear, logarithmic, and exponential. To minimize uncertainties arising from dense-matter physics, we adopt the realistic DDME2 EoS \citep{Xia:2022,Lalazissis:2005de}, which is consistent with nuclear and astrophysical constraints. Most importantly, for the first time in literature of $f(Q)$ gravity, we have employed the Bayesian inference framework to constrain the free model parameters and compare their predictions for NS properties with observational data from the NICER. This approach enables a systematic assessment of the viability of different 
$f(Q)$ models in the strong-field regime, highlighting the role of NSs as powerful probes of symmetric teleparallel gravity beyond cosmological scales. Therefore, our paper is organized as follows:

The organization of this paper is as follows. In Sec.~\ref{sec:2}, we briefly outline the geometrical formalism of $f(Q)$ gravity. The modified Tolman–Oppenheimer–Volkoff (TOV) equations governing the neutron-star structure in $f(Q)$ gravity are derived in Sec.~\ref{sec:3}. In Sec.~\ref{sec:4}, we introduce the three representative $f(Q)$ models considered in this study. The numerical methodology for solving the modified TOV equations and the Bayesian inference framework used for parameter estimation are presented in Sects.~\ref{sec:5} and \ref{sec:6}, respectively. Our results, including a comparative analysis of the different models, are discussed in Sec.~\ref{sec:7}. Finally, we summarize our findings and present concluding remarks in Sec.~\ref{sec:9}.

\section{Geometry of $f(Q)$ Gravity}\label{sec:2}

In the context of metric-affine geometry, a spacetime is characterized by the triplet $(\mathcal{M}, g_{\mu\nu}, \Gamma^\alpha_{~\mu\nu})$, where $\mathcal{M}$ denotes a four-dimensional differentiable manifold, $g_{\mu\nu}$ is the metric tensor (with signature $(-,+,+,+)$ or $(+,-,-,-)$, and $\Gamma^\alpha_{~\mu\nu}$ represents a general affine connection, independent of the metric. This framework allows for a more general geometric structure than that of standard Riemannian geometry, as it permits torsion and non-metricity in addition to curvature.

The affine connection defines a covariant derivative that governs how vectors and covectors change under parallel transport. Explicitly, for a vector $V^\alpha$ and a covector $V_\alpha$, the covariant derivatives are defined by:
\begin{align} 
\nabla_\mu V^\alpha &= \partial_\mu V^\alpha + \Gamma^\alpha_{~\mu\lambda} V^\lambda,\\
 \nabla_\mu V_\alpha &= \partial_\mu V_\alpha - \Gamma^\lambda_{~~\mu\alpha} V_\lambda. \end{align}



A general affine connection $\Gamma^\alpha_{~\mu\nu}$ can be decomposed into three independent components \citep{wang/2022}:
\begin{equation}
\Gamma^\alpha_{~\mu\nu} = \mathring{\Gamma}^\alpha_{~\mu\nu} + K^\alpha_{~\mu\nu} + L^\alpha_{~\mu\nu},
\end{equation}
where $\mathring{\Gamma}^\alpha_{~\mu\nu}$ is the Levi-Civita connection or widely known as the Christoffel symbol and defined by,
\begin{equation}
\mathring{\Gamma}^\alpha_{~\mu\nu} = \frac{1}{2} g^{\alpha\lambda} \left( \partial_\mu g_{\nu\lambda} + \partial_\nu g_{\mu\lambda} - \partial_\lambda g_{\mu\nu} \right).
\end{equation}
 Moreover, $K^\alpha_{~\mu\nu}$ is the contortion tensor associated with torsion tensor defined by
$T^\alpha_{~\mu\nu} := 2\Gamma^\alpha_{~[\mu\nu]}$
from which the contortion tensor is defined as,
\begin{equation}
K^\alpha_{~\mu\nu} = \frac{1}{2} \left( T^\alpha_{~\mu\nu} + T_{\mu\ \nu}^{\ \alpha} + T_{\nu\ \mu}^{\ \alpha} \right).
\end{equation}

Finally, the disformation tensor is given by
\begin{equation}
L^\alpha_{~\mu\nu} = \frac{1}{2} \left( Q^\alpha_{\ \mu\nu} - Q_{\mu\ \nu}^{\ \alpha} - Q_{\nu\ \mu}^{\ \alpha} \right),
\end{equation}
in terms of the non-metricity tensor
\begin{equation}
    Q_{\alpha\mu\nu} = \nabla_\alpha g_{\mu\nu}
    = \partial_\alpha g_{\mu\nu} - \Gamma^\beta_{\alpha\mu}g_{\beta\nu} - \Gamma^\beta_{\alpha\nu}g_{\mu\beta} .
\end{equation}

Symmetric teleparallel gravity (STG) describes a class of non-metric gravity theories in which both curvature and torsion of the affine connection vanish, so gravity is encoded purely by non-metricity, i.e.,
\begin{eqnarray}
   &&\hspace{0cm}  R^\lambda_{~~\mu\nu\sigma} =0 \quad \text{and} \quad T^{\alpha}_{~~\mu\nu} = 0 .
\end{eqnarray}

Therefore, in STG the connection reduces to
\begin{equation}
    \Gamma^{\alpha}_{~\mu\nu} = \left\{^{\,\alpha}_{\mu\nu}\right\} + L^{\alpha}_{~\mu\nu} .
\end{equation}

Next, we introduce the nonmetricity conjugate:
\begin{equation}
P^{\alpha\mu\nu} = -\frac{1}{4} Q^{\alpha\mu\nu} + \frac{1}{2} Q^{(\mu\ \nu)\alpha} + \frac{1}{4} \left( Q^\alpha - \tilde{Q}^\alpha \right) g^{\mu\nu}, 
\end{equation}
where its two independent traces are given by, 
\begin{eqnarray}
Q_\alpha = Q_{\alpha~~\mu}^{\, \, \, \mu},~ \text{and}~~ 
\tilde{Q}_\alpha = Q^\mu_{~~\alpha\mu}.
\end{eqnarray}
Finally, the nonmetricity scalar is defined as:
\begin{equation}
Q = -Q_{\alpha\mu\nu} P^{\alpha\mu\nu}. 
\end{equation}


Incorporating constraints via Lagrange multipliers, the modified Einstein-Hilbert action in $f(Q)$ gravity is given by~\citep{LH/2018}:
\begin{equation}
S = \int \sqrt{-g} ~ d^4x \left[ \frac{1}{2} f(Q) + \lambda_\alpha^{~~ \beta\mu\nu} R^\alpha_{~~ \beta\mu\nu} + \lambda^{~~\mu\nu}_{\alpha} T^{\alpha}_{~~ \mu\nu} + \mathcal{L}_m \right], \label{eq:action} \end{equation}
where $g$ denotes the determinant of the spacetime metric $g_{\mu\nu}$, $f(Q)$ is an arbitrary function of the nonmetricity scalar $Q$, $\lambda^\alpha_{\ \beta\mu\nu}$ and $\lambda^{~~~\mu\nu}_{\ \alpha}$ serve as Lagrange multipliers, and $\mathcal{L}_m$ is the Lagrangian density corresponding to matter.

Changing the action~\eqref{eq:action} with respect to the metric tensor, one arrives at the following field equation: \begin{eqnarray}
&&\hspace{0cm}-T_{\mu\nu} = \frac{2}{\sqrt{-g}} \nabla_\alpha \left( \sqrt{-g} f_Q P^\alpha_{\ \mu\nu} \right) + \frac{1}{2} g_{\mu\nu} f \nonumber
\\&&\hspace{0cm}+ f_Q \left( P_{\mu\alpha\beta} Q_\nu^{\ \alpha\beta} - 2 Q_{\alpha\beta\mu} P^{\alpha\beta}_{\ \ \nu} \right), \label{eq:field_eq}
\end{eqnarray} 

Here, the subscript $Q$ implies the differentiation of $f(Q)$ with respect to $Q$, that is, $f_Q \equiv \partial_Q f(Q)$. The standard definition of the energy-momentum tensor given by, \begin{equation}
T_{\mu\nu} \equiv -\frac{2}{\sqrt{-g}} \frac{\delta(\sqrt{-g} \mathcal{L}_m)}{\delta g^{\mu\nu}}. \label{eq:EM_tensor} 
\end{equation} 

As discussed in \citep{Jimenez:2018b}, the equation of motion given in Eq.(\ref{eq:field_eq}) can be rewritten  in a covariant form, similar to the standard Einstein gravitational field equations, as
\begin{equation}
f_Q \,\mathring{G}_{\mu\nu}
+ \frac{1}{2} g_{\mu\nu}\left(f_Q Q - f\right)
+ 2 f_{QQ} (\partial_\lambda Q)\, P^{\lambda}{}_{\mu\nu}
= T_{\mu\nu},\label{vv1}
\end{equation}
where \(\mathring{G}_{\mu\nu} = \mathring{R}_{\mu\nu} - \tfrac{1}{2}\mathring{R} g_{\mu\nu}\) denotes the Einstein tensor constructed from the Levi--Civita connection. Written in this form, the equation clarifies the role of diffeomorphism invariance in the theory. In the special case of STGR, where \(f(Q)=Q\), the left-hand side of Eq.~(\ref{vv1}) reduces identically to the Einstein tensor, which depends only on the metric. Consequently, the choice of affine connection does not influence the metric dynamics, in agreement with the discussion in the previous section. In contrast, for general \(f(Q)\) models with nonlinear dependence on \(Q\), the affine connection explicitly enters the field equations and therefore affects the evolution of the metric.

Upon variation of Eq.~\eqref{eq:action} with respect to the affine connection,  one can get:
\begin{equation} \nabla_\rho \lambda_{\alpha}^{~~~\nu\mu\rho} + \lambda^{~~\mu\nu}_{\alpha} = \sqrt{-g} f_Q P^\alpha_{~~ \mu\nu} + H_\alpha^{~~\mu\nu}, \label{eq:connection_var} \end{equation} 
where the hypermomentum tensor is defined as: 
\begin{equation} H_\alpha^{~~\mu\nu} = -\frac{1}{2} \frac{\delta \mathcal{L}_m}{\delta \Gamma^\alpha_{\ \mu\nu}}. \label{eq:hyper} \end{equation} Using the fact that the Lagrange multipliers are antisymmetric in $\mu$ and $\nu$, Eq.~\eqref{eq:connection_var} simplifies to:
\begin{equation} \nabla_\mu \nabla_\nu \left( \sqrt{-g} f_Q P^{\mu\nu}_{~~~\alpha} + H^{~~\mu\nu}_{\alpha} \right) = 0. \label{eq:antisym}
\end{equation} 

Assuming the hypermomentum contribution satisfies $\nabla_\mu \nabla_\nu H^{~~~\mu\nu}_{\alpha} = 0$ (for further explanation, readers may look into Ref.~\citep{LH/2018}), this reduces to:

\begin{equation} \nabla_\mu \nabla_\nu \left( \sqrt{-g} f_Q P^{\mu\nu}_{~~~~ \alpha} \right) = 0. \label{eq:conn_eq_simplified} \end{equation}

Next, in the absence of both curvature and torsion, the affine connection is expressible via~\citep{PhysRevD.98.044048}: 

\begin{equation}
\Gamma^\alpha_{\ \mu\nu} = \left( \frac{\partial x^\alpha}{\partial \xi^\lambda} \right) \partial_\mu \partial_\nu \xi^\lambda. \label{eq:inertial_conn} 
\end{equation} 

Furthermore, A convenient coordinate system called the coincident gauge may be adopted such that the affine connection vanishes identically, i.e., $\Gamma^\alpha_{\ \mu\nu} = 0$. Under this condition, the nonmetricity tensor becomes: \begin{equation} Q_{\alpha\mu\nu} = \partial_\alpha g_{\mu\nu}, \label{eq:coincident} \end{equation} greatly streamlining calculations by making the metric the sole dynamical variable. However, this choice restricts diffeomorphism invariance, an issue only circumvented in STGR~\citep{LH/2020}.

\section{Modified TOV in $f(Q)$ Gravity}\label{sec:3}

To describe the stellar structure, we consider the general form of a static, spherically symmetric spacetime given by the metric,  
\begin{align}
    \label{Eq: spherical-ansatz}
    ds^{2}=-e^{\nu(r)}dt^{2}+e^{\lambda(r)}dr^{2} + r^{2}d\Omega^{2}
    \,,
\end{align}
where \( d\Omega^2 = d\theta^2 + \sin^2\theta\,d\phi^2 \) represents the metric on the unit 2-sphere. $\nu(r)$ and $\lambda(r)$ represents the metric potential along the $-tt$ and $-rr$ direction. In this spherically symmetric case, we model the matter content as an isotropic fluid whose energy-momentum tensor takes the form of  
\begin{align}
    T_{\mu \nu}=\left( \varepsilon +P \right) u_{\mu}u_{\nu}+P g_{\mu \nu}
    \label{Eq: fluid}
    \,.
\end{align}
Here, \( u_\mu \) is the fluid's four-velocity, satisfying the orthogonality and normalization conditions:  
\( u^\mu u_\mu = -1 \), and \( u^\mu v_\mu = 0 \).  
The functions \( \varepsilon(r) \), \( P(r) \) denote the energy density and radial pressure of the fluid respectively.

From the modified field equations corresponding to the isotropic fluid and the metric above, the independent components of the field equations can be derived as:
\begin{eqnarray}
\label{Eq: eom-dens}
&&\hspace{-0.7cm}\varepsilon = \frac{f}{2} - f_Q\left( Q+\frac{1}{r^2}+\frac{e^{-\lambda}}{r}(\nu^{\prime}+\lambda^{\prime})\right)
\,, \\
\label{Eq: eom-presr}
&&\hspace{-0.7cm}P =- \frac{f}{2} + f_Q\left(Q+\frac{1}{r^2}\right)
\, , \\
\label{Eq: eom-prest}
&&\hspace{-0.7cm}P = -\frac{f}{2} + f_Q\Big\{ \frac{Q}{2}-e^{-\lambda}\big[\frac{\nu^{\prime\prime}}{2}+\big(\frac{\nu^{\prime}}{4}+\frac{1}{2r}\big) (\nu^{\prime}-\lambda^{\prime})\big]\Big\},~~~~~~~ \\
\label{Eq: eom-impos}
&&\hspace{0cm}0 = \frac{\cot\theta}{2}Q^{\prime}f_{QQ}
\,.
\end{eqnarray}

Using this line element in the context of symmetric teleparallel gravity, the nonmetricity scalar \( Q \) becomes a function of the radial coordinate and is given by  
\begin{align}
\label{Eq: spherical-nms}
Q(r) = -\frac{2e^{-\lambda}}{r}\left(\nu^{\prime}+\frac{1}{r}\right)
\,,
\end{align}
where the prime denotes a derivative with respect to \( r \).

Upon substituting the expression for \( Q \) from Eq.~(\ref{Eq: spherical-nms}) into the equations of motion Eq.~(\ref{Eq: eom-dens}) and Eq.~(\ref{Eq: eom-presr}), the modified version of the TOV equation in \( f(Q) \) gravity takes the form:
\begin{eqnarray}
&& P^{\prime}+\frac{\nu^{\prime}}{2}\left(\varepsilon+P\right) = f_{Q}\frac{e^{-\lambda}}{2r} \left[ \frac{2}{r}\left(\nu^{\prime}+\lambda^{\prime}\right) - \nu^{\prime}\left(\nu^{\prime}-\lambda^{\prime}\right) \right. \nonumber \\
&& \left. + \frac{4}{r^{2}}\left(1-e^{\lambda}\right) - 2\nu^{\prime\prime} \right] = 0.
\end{eqnarray}

By solving Eqs.~(\ref{Eq: eom-dens})--(\ref{Eq: eom-prest}) for a specific functional form of
$f(Q)$, together with appropriate boundary conditions, the metric functions
$\nu(r)$ and $\lambda(r)$ can be determined. As an illustrative case, for vacuum
solutions with $T_{\mu\nu}=0$, Eqs.~(\ref{Eq: eom-dens}) and (\ref{Eq: eom-presr})
imply the relation
\begin{equation}
\nu'(r) + \lambda'(r) = 0 .
\end{equation}

In general, Eqs.~(\ref{Eq: eom-dens})--(\ref{Eq: eom-prest}) can be reformulated as
a system of modified TOV equations in the framework of $f(Q)$ gravity. These equations describe the internal structure of neutron
stars and are supplemented by the continuity equation arising from the
conservation of the energy-momentum tensor $T_{\mu\nu}$, given by
\begin{eqnarray}
&&\hspace{0cm}P' + \frac{\nu'}{2}\,(\varepsilon + P) = 0, \\
&&\hspace{0cm}\lambda' = \frac{-k r (\rho + P)e^{\lambda} - \nu'}{f(Q)}, \\
&&\hspace{0cm}\nu'' = \frac{1}{2r f_Q}
\Big[
2rQf_Q e^{\lambda}
- (r\nu' + 2)(\nu' - \lambda')f_Q  \nonumber \\
&& \hspace{1.2cm}
- 2\bigl(2kP + f(Q)\bigr) r e^{\lambda}
\Big].
\end{eqnarray}

By specifying an EoS and providing suitable initial conditions
for the metric functions $\nu(r)$ and $\lambda(r)$, as well as for the energy
density $\varepsilon$ and pressure $p$, the stellar structure of NSs can be fully determined in
$f(Q)$ gravity. In the particular limit $f(Q)=-Q$,
Eqs.~(\ref{Eq: eom-dens})--(\ref{Eq: eom-prest}) reduce to the standard field
equations of GR. These equations therefore allow for the computation of NS
configurations for various $f(Q)$ gravity models.

\section{ $f(Q)$ Models}\label{sec:4}
In this work, we investigate the internal structure of NSs in the context of 
$f(Q)$ gravity by considering three representative functional forms, namely linear, logarithmic, and exponential models. These formulations extend the standard gravitational action through both linear and nonlinear dependencies on the nonmetricity scalar 
$Q$, leading to modified field equations that govern stellar equilibrium. A systematic examination of these models allows us to assess how deviations from GR influence the macroscopic characteristics of NSs.
\subsection{Linear Model}
To investigate the role of different functional forms of $f(Q)$, we first focus on the linear model \citep{wang/2022},
\begin{eqnarray}
f(Q)=Q_0 \bigg[\alpha+\beta \Big( \frac{Q}{Q_0} \Big)  \bigg],
\end{eqnarray}
which emerges directly from Eq.~(\ref{Eq: eom-impos}) upon imposing the condition $f_{QQ}=0$. This constraint restricts the gravitational action to depend linearly on the nonmetricity scalar $Q$. The choice of the linear form of the model is primarily motivated by dimensional consistency and its theoretical analogy with GR. In the Einstein-Hilbert action, the gravitational Lagrangian density is proportional to the Ricci scalar $R$. In STEGR, the Ricci scalar is replaced by a function of the non-metricity scalar, $f(Q)$. Therefore, in order to maintain dimensional consistency of the gravitational action, the function $f(Q)$ must carry the same physical dimension as $R$, namely $[\mathrm{length}]^{-2}$.
In a static spherically symmetric spacetime, the non-metricity scalar $Q$ also has the dimension $[\mathrm{length}]^{-2}$. To construct generalized functional forms such as logarithmic and exponential corrections, we introduce a constant scale $Q_0$ with the same dimension as $Q$. This guarantees that the ratio $Q/Q_0$ is dimensionless, ensuring that functions such as $\ln(Q/Q_0)$ or $\exp(Q/Q_0)$ are mathematically well defined.
Consequently, the parameters $\alpha$ and $\beta$ are dimensionless constants that quantify deviations from GR, while $Q_0$ sets the characteristic non-metricity scale of the theory. The adopted linear form is therefore not arbitrary; rather, it is motivated by dimensional consistency, structural coherence with the Einstein-Hilbert framework, and uniformity with the logarithmic and exponential models considered in this work.

The linear $f(Q)$ model constitutes the minimal extension of STG and is exactly reduced to GR for $\alpha=0$ and $\beta = -1$. Owing to the absence of higher-order derivatives, the resulting field equations remain second order, ensuring mathematical consistency and freedom from additional dynamical degrees of freedom. Consequently, this model provides a well-controlled theoretical baseline for isolating leading-order gravitational modifications and serves as a natural reference against which nonlinear $f(Q)$ models can be systematically assessed in the context of NS structure.

\subsection{Logarithmic Model}

We now turn to a logarithmic form of the $f(Q)$ function, which introduces scale-dependent corrections to the gravitational action and has been widely studied in \citep{log}. The model is specified as
\begin{equation}
f(Q) = -Q + \alpha \ln\!\left[ \beta \left( \frac{Q}{Q_0} \right) \right],
\end{equation}
where $\alpha$ and $\beta$ are constant parameters controlling the magnitude and scale of the logarithmic correction, and $Q_0$ is a constant with dimensions of $[\mathrm{length}]^{-2}$ introduced to ensure a dimensionless argument of the logarithm. In the limit $\alpha \rightarrow 0$, the logarithmic contribution vanishes and the theory is reduced to the symmetric teleparallel formulation equivalent to GR.

The logarithmic $f(Q)$ model is motivated by its ability to introduce controlled scale-dependent deviations from GR. Logarithmic corrections naturally emerge in effective field theories and quantum-inspired gravity \citep{eft}, encoding running gravitational effects that remain negligible in the weak-field limit while becoming relevant in intermediate and strong-field regimes, such as those encountered in compact objects.

From a cosmological standpoint, logarithmic $f(Q)$ models have been widely applied to study late-time cosmic acceleration, dynamical dark energy, and phase-space stability \citep{log1}. These properties provide a compelling motivation to investigate their implications for NS interiors, where strong gravity may amplify scale-dependent effects.

\subsection{Exponential Model}

Finally, we consider an exponential form of the $f(Q)$ function, which introduces nonlinear corrections to the gravitational action and has been widely explored in the literature \citep{simu}. The model is defined as
\begin{equation}
f(Q) = -Q + \alpha Q_0 \left[ 1 - \exp\!\left(-\beta \sqrt{\frac{Q}{Q_0}} \right) \right],
\end{equation}
where $\alpha$ and $\beta$ are dimensionless parameters characterizing the strength and scale of deviations from GR, and $Q_0$ is a constant with dimensions of $[\mathrm{length}]^{-2}$ introduced to render the argument of the exponential dimensionless. In the limit $\beta \rightarrow 0$, the exponential correction vanishes and the theory reduces to the symmetric teleparallel formulation equivalent to GR.

The exponential model is physically motivated by its ability to generate smooth and controlled deviations from GR while remaining well behaved in both the weak and strong field regimes \citep{gaurav}. The exponential suppression ensures that nonlinear corrections become significant only above a characteristic scale set by $Q_0$, allowing the theory to recover GR at low curvatures while permitting appreciable modifications in high-density environments. Such properties make this model particularly attractive for studying compact objects, where strong gravitational fields can amplify deviations from standard gravity.

Beyond astrophysical applications, exponential $f(Q)$ models have been extensively employed in cosmological contexts, including inflationary dynamics, constraints from big bang nucleosynthesis, and phase-space analyses of cosmic evolution \citep{bbn}. These successes motivate their application to NS interiors, where they provide a theoretically consistent framework to probe the impact of nonlinear nonmetricity effects on stellar structure.

\section{Boundary Conditions and Equation of State}\label{sec:5}

For each $f(Q)$ model, we compute the interior and exterior structure of neutron stars by numerically integrating the modified TOV equations. The system consists of coupled ordinary differential equations for the enclosed mass $m(r)$, the metric functions $\lambda(r)$ and $\nu(r)$, and the pressure $P(r)$. These equations are solved using a fourth-order Runge-Kutta (RK4) scheme.

The integration is initiated at the stellar center, where regularity imposes the boundary conditions
\begin{eqnarray}
m(0) = 0, \qquad \lambda(0) = 0, \qquad \nu'(0) = 0 .
\end{eqnarray}
The central pressure $P_c$ is determined from a realistic tabulated equation of state (EoS). In this work, we adopt the density-dependent relativistic mean-field DDME2 EoS \citep{Lalazissis:2005de}, which provides a consistent relation between pressure and energy density. The central energy density $\varepsilon_c$ is varied over the range $100$--$1700~\mathrm{MeV\,fm^{-3}}$, corresponding to several times the nuclear saturation density.

For a given $\varepsilon_c$, the modified TOV equations are integrated radially outward with a fixed step size $\Delta r = 0.01~\mathrm{km}$ until the pressure vanishes, $P(R)=0$, which defines the stellar surface at radius $R$. The total gravitational mass is then obtained as $M = m(R)$. To ensure a physically consistent global solution, the central value of the metric function $\nu(r_0)$ is determined using a shooting method: the trial value of $\nu(r_0)$ is iteratively adjusted until the interior solution matches smoothly to the exterior Schwarzschild spacetime at the surface, satisfying
$e^{\nu(R)} = e^{-\lambda(R)}$ .
Repeating this procedure over the full range of central energy densities yields the mass--radius relation for each $f(Q)$ model. This allows us to systematically assess the impact of nonmetricity-induced corrections on neutron-star structure and to compare the resulting configurations with observational constraints.

\section{Bayesian Estimation}\label{sec:6}
A Bayesian approach allows one to perform a detailed statistical analysis of the parameters of a model for a given set of data. This approach is based on \textit{Bayes' theorem}. It combines prior knowledge with fit data to produce the \textit{posterior probability} distribution for model parameters. This posterior distribution effectively quantifies the probability of every possible parameter hypothesis. From this, we can study individual parameter distributions and examine correlations between them. For a given dataset $\mathcal{D}$ and a hypothesis $H(\theta)$, \textit{Bayes' theorem} is expressed as
\[
P(\theta | \mathcal{D}, H) = \frac{{\mathcal{L}(\mathcal{D} | \theta, H) P(\theta | H)}}{P(\mathcal{D} | H)}.
\]
The components of \textit{Bayes' theorem} are defined as follows:
\begin{itemize}
    \item \textbf{Posterior probability \( P(\theta \mid \mathcal{D}, H) \): }The updated probability of parameters \(\theta\) after observing data \(\mathcal{D}\) within the model \(H\).
    \item \textbf{Likelihood \( \mathcal{L}(\mathcal{D} \mid \theta, H) \):} The probability of observing the data \(\mathcal{D}\) given specific parameter values \(\theta\) under the hypothesis \(H\).
    \item \textbf{Prior probability \( P(\theta \mid H)\):} The initial probability assigned to \(\theta\) before seeing the data, based on existing knowledge or assumptions.
    \item \textbf{Marginal likelihood (evidence) \( P(\mathcal{D} \mid H) \):} A normalizing constant representing the probability of the data \(\mathcal{D}\) under \(H\), averaged over all possible values of \(\theta\).
\end{itemize}
\subsection{Likelihood} 
In Bayesian analysis, the likelihood function quantifies the probability of the observed data for a given set of model parameters  $\theta$. It can be computed for a given posterior distributions from astrophysical observations as follows:

\textbf{X-ray observations (NICER):} 
X-ray observations of NS hot spots by the NICER mission have given some of the strongest constraints on NS masses and radii. The X-ray measurements provide the mass and radius of individual pulsars, both isolated and in binaries. These measurements offer a direct way to study the dense-matter EoS.

The main NICER results come from two key pulsars. For PSR J0030+0451 (at a canonical mass), mass–radius posteriors were published independently by two groups:~\citep{Riley:2019yda}~\footnote{~\href{https://zenodo.org/records/8239000}{https://zenodo.org/records/8239000}}  and ~\citep{Miller:2019cac}~\footnote{~\href{https://zenodo.org/record/3473466}{https://zenodo.org/record/3473466}}. For PSR J0740+6620 (near the maximum mass), similar posteriors were released by~\citep{Riley:2021}~\footnote{~\href{https://zenodo.org/records/4697625}{https://zenodo.org/records/4697625}} and ~\citep{Miller:2021}~\footnote{~\href{https://zenodo.org/records/4670689}{https://zenodo.org/records/4670689}}.  Recently, NICER has provided additional mass–radius measurements for other pulsars such as PSRJ0437+4715 from~\citep{Choudhury:2024xbk}~\footnote{~\href{https://zenodo.org/records/13766753}{https://zenodo.org/records/13766753}} and PSRJ0614+3329 from~\citep{Mauviard:2025dmd}~\footnote{~\href{https://zenodo.org/records/15603406}{https://zenodo.org/records/15603406}}. 

For the NICER mass-radius observations, the likelihood is formally expressed by marginalizing over the stellar mass:
\begin{eqnarray}
&&\hspace{0cm}\mathcal{L}^{\rm NICER}(\boldsymbol{\theta}_{\rm f(Q)})
\equiv P\!\left(d_{\rm X\text{-}ray}\mid \boldsymbol{\theta}_{\rm f(Q)}\right) \nonumber \\
&&\hspace{0cm}= \int_{M_l}^{M_u} dm \;
P\!\left(m \mid \boldsymbol{\theta}_{\rm f(Q)}\right)
P\!\left(d_{\rm X\text{-}ray} \mid m, R\!\left(m, \boldsymbol{\theta}_{\rm f(Q)}\right)\right).~~~~~~~
\end{eqnarray}

Here, $M_l$ denotes the lower mass bound, taken to be $1\,M_\odot$, and
$M_u(\boldsymbol{\theta}_{\rm f(Q)})$ represents the maximum NS
mass obtained by solving the modified TOV
equations \citep{Oppenheimer:1939ne, Tolman:1939jz} for a given set of $f(Q)$ model parameters
$\boldsymbol{\theta}_{\rm f(Q)}$, assuming the DDME2 EoS.

Suppose there are three statistically independent datasets $A$, $B$, and $C$, with corresponding Bayesian likelihoods
$\mathcal{L}^A(\boldsymbol{\theta})$,
$\mathcal{L}^B(\boldsymbol{\theta})$, and
$\mathcal{L}^C(\boldsymbol{\theta})$,
where $\boldsymbol{\theta}$ denotes the model parameters.
Under the assumption of statistical independence, the joint likelihood
is given by
\begin{equation}
\mathcal{L}(\boldsymbol{\theta})
= \prod_{i=A,B,C} \mathcal{L}^i(\boldsymbol{\theta}) .
\end{equation}

\subsection{Bayes Factor} \label{bayesfactor}
A \textit{Bayes factor} (BF) measures the strength of evidence provided by the data in favor of one hypothesis \( H_1 \) over another \( H_2 \).
It is defined as:
\begin{equation}
\text{BF}_{12} = \frac{P(D | H_1)}{P(D | H_2)},
\end{equation}
where,
\begin{itemize}
    \item \( P(D | H_1) \) is the likelihood of the data \( D \) under hypothesis \( H_1 \).
    \item \( P(D | H_2) \) is the likelihood of the data \( D \) under hypothesis \( H_2 \).
\end{itemize}
\subsubsection{Log of the Bayes Factor }
Taking the \textit{logarithm} of the Bayes factor provides a more practical scale for interpretation and calculation \citep{Imam:2025lut}. The log Bayes factor is:
\begin{equation}
    \ln(\text{BF}_{12}) = \ln\left(\frac{P(D | H_1)}{P(D | H_2)}\right).\label{logBayes}
\end{equation}
This transformation provides a more practical scale for quantifying the strength of evidence that favors one hypothesis.
\subsubsection{Interpretation of the Bayes Factor }\label{IntpBF}

To assess the strength of evidence, the values of the Bayes factor are often assessed using the following conventional scale \citep{Kass01061995, Jarosz2014}:
\begin{itemize}
    \item \( \ln(\text{BF}) > 1 \): Strong evidence for hypothesis \( H_1 \).
    \item \( 0 < \ln(\text{BF}) < 1 \): Weak to moderate evidence for hypothesis \( H_1 \).
    \item \( \ln(\text{BF}) = 0 \): No evidence for either hypothesis.
    \item \( -1 < \ln(\text{BF}) < 0 \): Weak to moderate evidence for hypothesis \( H_2 \).
    \item \( \ln(\text{BF}) < -1 \): Strong evidence for hypothesis \( H_2 \).
\end{itemize}

\section{Results}\label{sec:7}
\subsection{Priors}
\begin{figure*}[!ht]
    \centering
    \includegraphics[height=5.8cm,width=\linewidth]{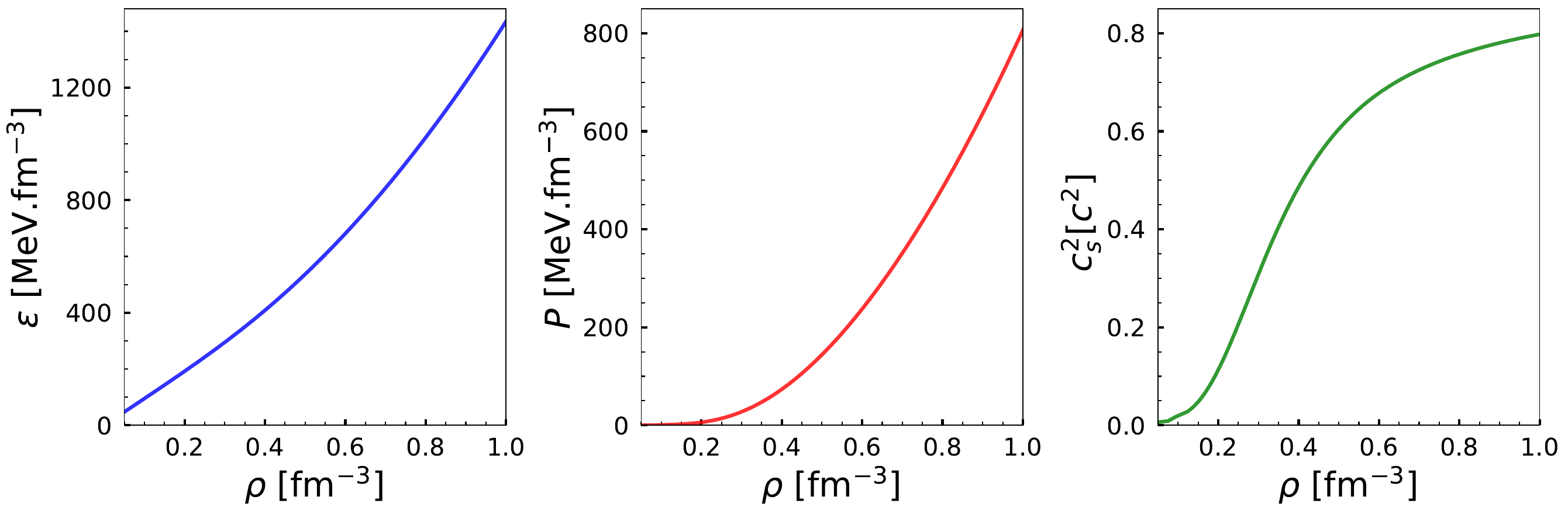}
    \caption{The energy density ($\epsilon$) (in MeV.fm$^{-3}$)(left panel), pressure ($P$) (in MeV.fm$^{-3}$) (middle panel), and speed of sound ($c_s^2$) (in $c^2$) (right panel) as a function of baryon density ($\rho$) (in fm$^{-3}$) for DDME2.}
    \label{fig1}
\end{figure*}

We investigate the impact of observational data on NS properties mass, radius, tidal deformability, and compactness using a set of $f(Q)$ models: the linear, logarithmic, and exponential forms detailed in Sec. \ref{sec:4}. These models are constrained within a Bayesian framework using NICER mass–radius posteriors. To assess the flexibility of each $f(Q)$ model, we adopt the DDME2 hadronic EoS as a baseline. In Fig.~\ref{fig1}, we plot the energy density ($\epsilon$)(left), pressure ($P$) (middle), and speed of sound ($c_s^2$) (right) as a function of baryon density $\rho$ for DDME2. The prior distributions for the model parameters are listed in table \ref{tab1}. Since these specific models have not previously been explored in the context of astrophysical objects, we have adopted sufficiently broad prior ranges for the parameters $\alpha,\beta,Q_0$. The chosen intervals are as wide as possible while still ensuring the existence of physically stable solutions.

\begin{table}
\caption{\label{tab1}Uniform priors on the $f(Q)$ model parameters.}
\centering
\begin{tabular}{lc}
\hline\hline
Parameter & Prior \\
\midrule
$\alpha$ & $[-15,15]$ \\
$\beta$  & $[-3,5]$ \\
$Q_0$    & $[-1,1]$ \\
\hline
\hline
\end{tabular}
\end{table}

\subsection{Posterior distribution of $f(Q)$ model parameters and NS Properties}

\begin{figure*}
    \centering
    \includegraphics[height=13.6cm,width=12.0cm]{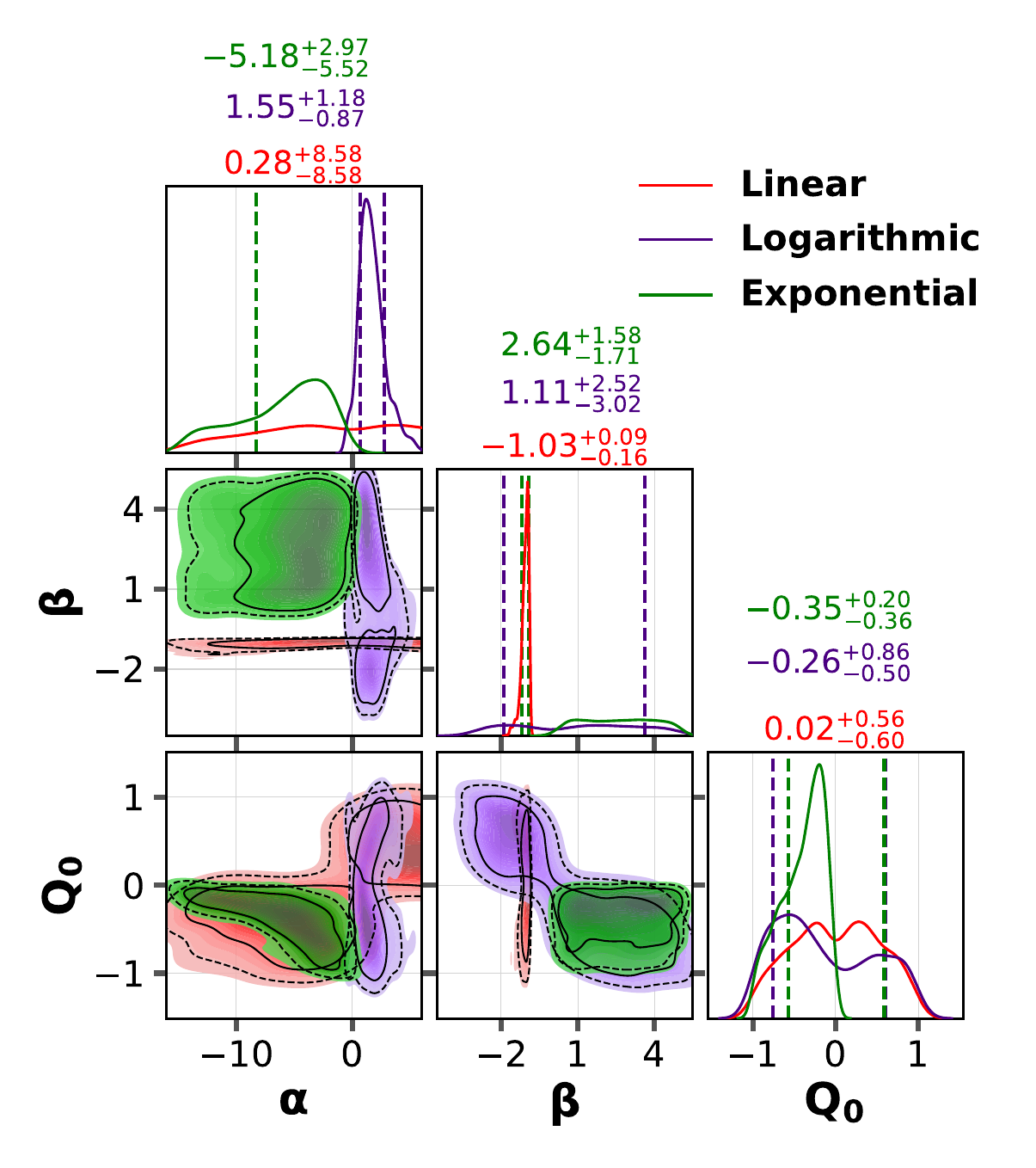}
    \caption{The marginalized posterior distributions of the
$f(Q)$ model parameters, obtained through Bayesian inference, for linear  (red), logarithmic  (purple), and exponential  (green) models. The vertical lines indicate
the 68\% confidence interval of the parameters. The confidence ellipses
for two-dimensional posterior distributions are plotted with 1$\sigma$, 2$\sigma$ and 3$\sigma$confidence intervals.}
    \label{fig2}
\end{figure*}

The corner plots for the marginalized posterior distributions for $f(Q)$ model parameters of linear (red), logarithmic (purple), and exponential (green) models are shown in Fig.~\ref{fig2}. The marginalized one-dimensional posterior distributions for the parameters are displayed along the diagonals of the corner plots. The median values of the parameters and their uncertainties $2\sigma$, drawn from the marginalized posterior distributions, are shown. The vertical lines indicate the 68\% confidence interval of the parameters. We also plot the 2D confidence ellipses along with the off-diagonal corner plots, corresponding to the credible intervals $1\sigma$, $2\sigma$, and $3\sigma$. The shapes and orientations of these ellipses indicate the presence or absence of correlations between the parameters.

\begin{table}
\caption{\label{tab2}The median values and associated 68\%(90\%) uncertainties for the parameters from their marginalized posterior
distributions. The results are obtained for linear, logarithmic and exponential models. } 
\renewcommand{\arraystretch}{1.4}
  
  \begin{tabular}{cccc}
  \hline\hline
\multirow{2}{*}{Models} & \multicolumn{3}{c}{Parameters} \\[1.0ex]
\cline{2-4}
                       & $\alpha$ & $\beta$ & $Q_0$ \\ \hline
 \hline
 Linear & $0.28^{+8.58(12.45)}_{-8.58(-12.84)}$ & $-1.03^{0.09(0.13)}_{-0.16(-0.34)}$ & $0.022^{+0.56(0.81)}_{-0.60(-0.88)}$ \\ [1.3ex]
 Logarithmic & $1.55^{+1.18(2.41)}_{-0.87(1.33)}$ & $1.11^{+2.52(3.45)}_{-3.02(3.77)}$ & $-0.26^{+0.86(1.12)}_{-0.50(0.66)}$\\ [1.3ex]
 Exponential & $-5.18^{+2.97(3.89)}_{-5.52(8.31)}$ & $2.64^{+1.58(2.15)}_{-1.71(2.14)}$ & $-0.35^{+0.20(0.28)}_{-0.36(0.54)}$\\ [1.3ex]
 \hline \hline
  \end{tabular}
   
\end{table}

\begin{table*}
\caption{\label{tab3}The median values and associated 68\%(90\%) uncertainties for the NS properties, namely the tidal deformability ($\Lambda_{1.4}$), radii ($R_{1.4}$ and $R_{2.07}$ and maximum
mass ($M_{max}$) from their posterior distribution.} 
  \centering

  \begin{tabular}{ccccc}
  \hline\hline
\multirow{2}{*}{Models} & \multicolumn{4}{c}{NS Properties} \\[1.0ex]
\cline{2-5}
                       & $\Lambda_{1.4}$& $R_{1.4}$ & $R_{2.07}$& $M_{max}$\\ \hline
 \hline
 Linear & $143.99^{+86.30(166.57)}_{-38.62(-53.07)}$ & $11.19^{+0.53(0.94)}_{-0.38(-0.57)}$ & $11.68^{+0.39(0.73)}_{-0.40(-0.93)}$ &$2.30^{+0.34(0.54)}_{-0.20(-0.29)}$\\ [1.3ex]
 Logarithmic & $135.82^{+64.59(179.65)}_{-23.33(32.21)}$ & $11.09^{+0.48(1.05)}_{-0.22(0.32)}$ & $12.08^{+0.36(0.65)}_{-0.25(0.42)}$ &$2.36^{+0.28(0.42)}_{-0.23(0.32)}$\\ [1.3ex]
 Exponential & $156.95^{+84.02(177.68)}_{-41.73(55.77)}$ & $11.27^{+0.53(0.97)}_{-0.36(0.52)}$ & $11.76^{+0.50(0.83)}_{-0.36(0.50)}$ &$2.29^{+0.11(0.16)}_{-0.13(0.21)}$\\ [1.3ex] \hline \hline
  \end{tabular}
 
\end{table*}
The correlation between $\alpha$ and $Q_0$ is moderate in the linear case ($r \sim 0.64$) and in the exponential case ($r \sim -0.64$) and negligible in the logarithmic case ($r \sim -0.14$). A strong correlation between $\beta$ and $Q_0$ is found only for the logarithmic model ($r \sim -0.79$). The median values of the model parameters and the corresponding confidence intervals of 68\% (90\%) obtained from the
marginalized posterior distributions are listed in Table~\ref{tab2}. The Bayesian analysis indicates that the constraining power of the data depends sensitively on how each parameter modifies the effective gravitational sector in the three $f(Q)$ parameterizations. In the linear model, the parameter 
$\beta = -1.03^{+0.09}_{-0.16}$ 
 is tightly constrained, reflecting its direct role in determining the effective linear coupling to $Q$. Since the GR limit corresponds to a specific linear scaling, the narrow credible interval indicates that the data strongly restrict deviations from this behavior. 
For the logarithmic model, the amplitude parameter 
$\alpha = 1.55^{+1.18}_{-0.87}$ 
is comparatively well constrained, indicating that the strength of the logarithmic correction is observationally relevant. However, 
$\beta = 1.11^{+2.52}_{-3.02}$ 
exhibits large asymmetric uncertainties, signaling degeneracy between the amplitude and scale of the modification and allowing partial compensation that weakens individual constraints. The scale parameter 
$Q_0 = -0.26^{+0.86}_{-0.50}$ 
is only moderately constrained, suggesting limited sensitivity of the data to the precise normalization inside the logarithmic term.
In the exponential model, the strongest constraint is obtained for 
$Q_0 = -0.35^{+0.20}_{-0.36}$, 
indicating that the characteristic scale governing the exponential suppression of deviations from GR is well determined. The parameter 
$\beta = 2.64^{+1.58}_{-1.71}$, 
 which controls the rate at which the model departs from the GR limit, is moderately constrained. 
Overall, parameters that directly control the effective linear coupling or the characteristic transition scale away from the GR limit are more tightly constrained, whereas additive or amplitude like parameters remain weakly determined due to degeneracies in the background expansion behavior.

\begin{figure*}
    \centering
    \includegraphics[height=12cm,width=12cm]{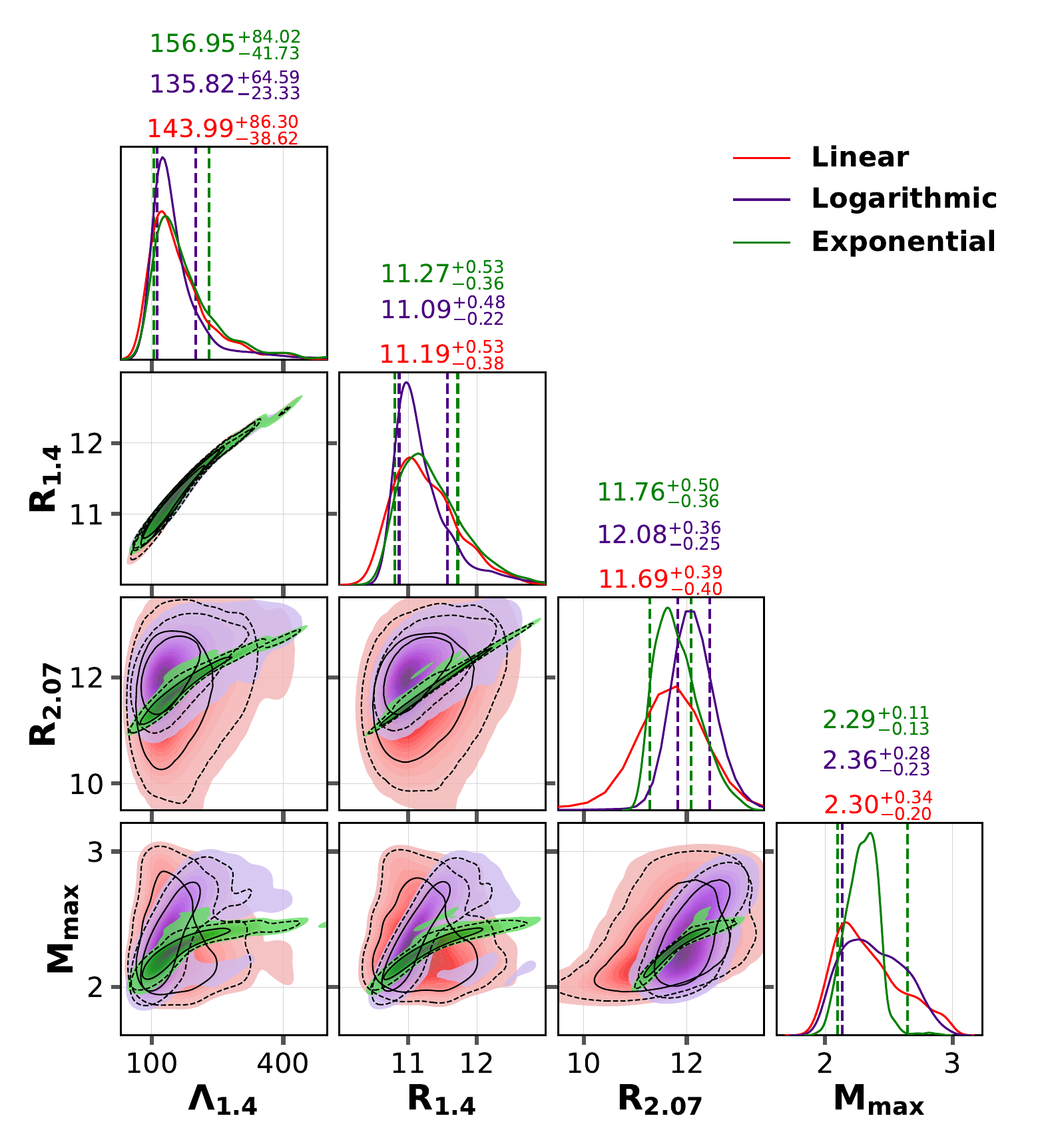}
    \caption{Corner plots for the marginalized posterior distributions
(salmon) of the tidal deformability $\Lambda_{1.4}$, radii $R_{1.4}$ (km) and $R_{2.07}$ (km) and the maximum mass $M_{max}$ ($M_\odot$) for linear (red), logarithmic (purple), and exponential (green). }
    \label{fig3}
\end{figure*}

We have obtained the distributions of NS properties such as tidal deformability $\Lambda_{1.4}$, radii $R_{1.4}$ (km) and $R_{2.07}$ (km) and the maximum mass $M_{max}$ ($M_\odot$)  using the posterior distributions for the parameters corresponding to the linear, logarithmic and exponential models as listed in table~\ref{tab2}. The corner plots for these NS properties are shown in Fig.~\ref{fig3}. The same as Fig.\ref{fig2}, the vertical lines, median values with $2\sigma$ uncertainties, and 2D confidence ellipses along the off-diagonal line are shown. It is clear from the off-diagonal graphs that $\Lambda_{1.4}$ is strongly correlated with $R_{1.4}$ for all models ($r\sim 0.96$). In the linear model, the maximum mass $M_{\max}$ is found to be statistically uncorrelated with the other neutron star observables, as indicated by nearly circular and non-tilted confidence contours. This absence of correlation implies that variations in the maximum mass occur independently of changes in other stellar properties, suggesting that the linear modification does not introduce significant structural coupling in the equilibrium configuration.
In contrast, the logarithmic model exhibits a mild but noticeable correlation between $M_{\max}$ and other neutron star properties, reflected by moderately tilted confidence contours. This indicates that the logarithmic correction induces a partial coupling between the maximum mass and the internal stellar structure, allowing variations in one quantity to influence the other within the allowed parameter space.
The exponential model, however, demonstrates a strong and pronounced correlation between $M_{\max}$ and the remaining neutron star properties, characterized by highly elongated and tilted contours. This behavior signifies a substantial structural interdependence introduced by the exponential modification, where changes in the gravitational parameters directly impact the global stellar configuration. Consequently, the maximum mass in this case is highly sensitive to the underlying modified gravity parameters, revealing a stronger departure from the effective GR-like behavior compared to the linear and logarithmic scenarios.
  We have summarized the median values of the properties of NS together with 68\% (90\%) confidence intervals in table~\ref{tab3}. It shows how different models for $f(Q)$ theory affect these astrophysical quantities.
The logarithmic model predicts the highest maximum mass ($2.36_{-0.23}^{+0.28}$) while the linear and exponential model shows the widest uncertainty ranges for most properties, particularly for $\Lambda_{1.4}$, indicating different constraint sensitivities. The radii predictions for $R_{1.4}$ and $R_{2.07}$ are remarkably similar for all models , differing by less than $\sim$ 0.4 km in the median values. This suggests that within these $f(Q)$ models, the predicted radius does not change significantly with mass in this range.

\begin{figure*}
    \centering
    \includegraphics[width=\linewidth]{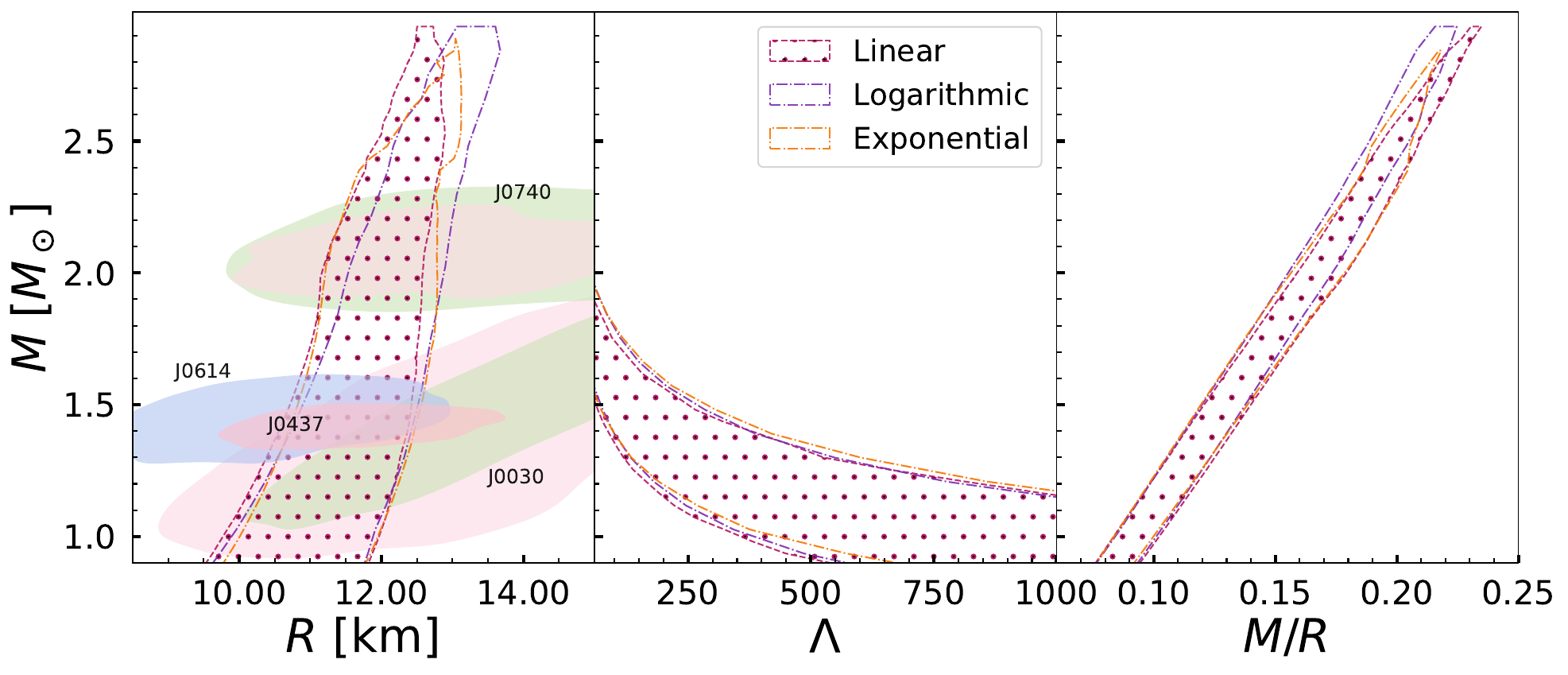}
    \caption{The 95\% confidence interval distributions for the radius $R$ (km) (left panel), tidal deformability $\Lambda$ (middle panel) and compactness $M/R$ (right panel) as a function of NS mass $M$ ($M_\odot$). The astrophysical observations incorporated in the Bayesian framework are also shown.}
    \label{fig4}
\end{figure*}

\begin{figure*}
    \centering
    \includegraphics[width=\linewidth]{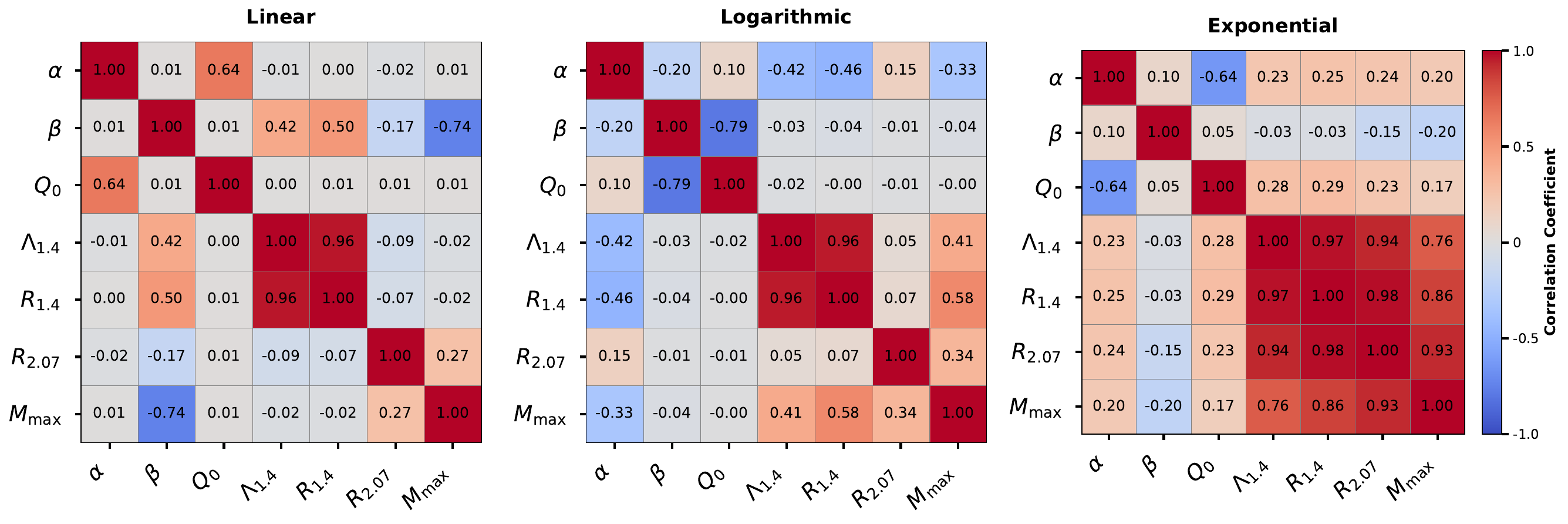}
    \caption{The Pearson's correlation coefficients among parameters and selected NS properties for linear (left panel), logarithmic (middle panel), and exponential(right panel).}
    \label{fig5}
\end{figure*}

Figure~\ref{fig4} displays the 95\% confidence bands for the $M$–$R$ (left panel), $M$–$\Lambda$ (middle panel), and $M$–$M/R$ (right panel) relations for the linear (magenta), logarithmic (purple), and exponential (orange) models. All NICER observational constraints incorporated in the Bayesian framework are also shown.
The $M-R$ curves for all models pass through the centers of the observational bands, indicating that the Bayesian inference successfully reproduces the observed data. The exponential model exhibits the widest confidence band, particularly in the 1–2 $M_\odot$ mass range, consistent with its larger parameter uncertainties (see Table~\ref{tab3}). In contrast, the linear model shows the narrowest band, reflecting its tighter parameter constraints. The logarithmic model yields broader distributions above $2 M_\odot$, suggesting a greater capacity to produce stiffer high-mass $M$–$R$ configurations. 
A key result of our analysis is that all three constrained $f(Q)$ models predict a maximum neutron star mass exceeding $2.5\,M_{\odot}$. This places the stable configurations within the lower mass gap region ($\sim 2.5$--$5\,M_{\odot}$), where the nature of compact objects remains uncertain. The $95\%$ confidence bands of the mass--radius relation consistently extend beyond $2.5\,M_{\odot}$ and reaching nearly about $3~M_{\odot}$, indicating that the Bayesian-constrained $f(Q)$ gravity parameters allow significantly more massive stable stars than typically obtained in standard GR scenarios. The appearance of mass-gap configurations therefore emerges naturally from the constrained parameter space and constitutes a clear, testable prediction of the present $f(Q)$ framework.
For the $M-\Lambda$ distributions, all three models follow similar trends. In terms of compactness, the linear model reaches higher $M/R$ values than the others at a given mass. The different compactness trends between models highlight how each $f(Q)$ functional form alters the mass–radius relation, particularly in the high-density regime.\\
The Pearson's correlation coefficients among all \( f(Q) \) model parameters (\( \alpha \), \( \beta \), and \( Q_0 \)) and the NS properties (\( \Lambda_{1.4} \), \( R_{1.4} \), \( R_{2.07} \), and \( M_{\text{max}} \)) are shown in Fig.~\ref{fig5}. The left, middle, and right panels correspond to the linear, logarithmic, and exponential models, respectively. In the linear model, \( \beta \) is strongly correlated with the $M_{\text{max}}$, radius and tidal deformability at the canonical mass, with correlation coefficients of \( -0.74\), \(-0.79 \) and \( -0.70 \), respectively. This indicates that \( \beta \) plays a key role in defining the canonical NS properties in the linear case. In the logarithmic model, \( \alpha \) is moderately correlated with these canonical properties. In contrast, for the exponential model, no strong correlation is found between individual parameters and NS properties, suggesting that the parameters \textbf{collectively} relate to NS properties.

\subsection{Model Comparison}

\begin{figure*}
    \centering
    \includegraphics[width=\linewidth]{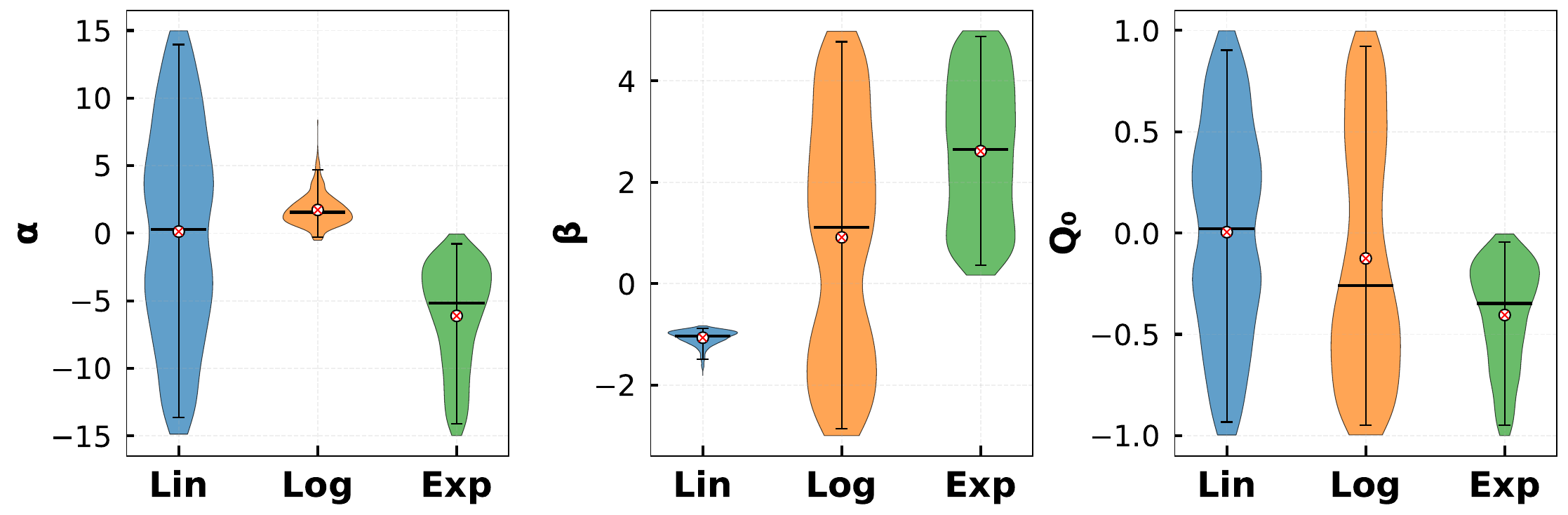}
    \caption{Violin plots of the posterior distributions for the \( f(Q) \) model parameters \( \alpha \) (left), \( \beta \) (middle), and \( Q_0 \) (right) for the linear (Lin), logarithmic (Log), and exponential (Exp) models.}
    \label{fig6}
\end{figure*}

\begin{table}
\caption{The logarithmic Bayes factor for all $f(Q)$ models considered for this analysis. Here we consider exponential as \( H_2 \) hypothesis always and other models as \( H_1 \) to compute \(\ln(\text{BF}_{12})\) using Eq. (\ref{logBayes}).\label{tab4}}
\centering
\begin{tabular}{cc}
\hline\hline
\textbf{Model} & \textbf{\(\ln(\text{BF}_{12})\) } \\
\hline
Linear & -1.81 \\
Logarithmic & -0.43\\
\hline \hline
\end{tabular}
\end{table}

Since we use a large number of $f(Q)$ models for this analysis, it would be good to compare these models statistically using the
Bayes factor as discussed in Sec.\ref{bayesfactor}. In Table~\ref{tab4}, we listed the logarithmic values of the Bayes factor computed using Eq. (\ref{logBayes}) for all models. The interpretation of the Bayes factor is discussed in Sec.\ref{IntpBF}. We consider the exponential model as hypothesis \( H_2 \) (because of large negative values of log evidence) and other models as \( H_1 \) to compute \(\ln(\mathrm{BF}_{12}) \). From the large negative values of $\ln(\mathrm{BF}_{12})$, it is clear that exponential is definitely the best model in the Bayesian inference.  The Bayes factors computed for all models consistently rank the models in the same order: exponential as the best, followed by logarithmic, then linear.

Violin plots provide a detailed view of posterior distributions by merging a box plot’s summary statistics with a smooth density curve. This allows for easy visual comparison of features such as skewness, multimodality, and the overall spread of model parameter constraints.
Figure~\ref{fig6} shows the Violin plots of the posterior distributions for the \( f(Q) \) model parameters \( \alpha \) (left), \( \beta \) (middle), and \( Q_0 \) (right) for the linear (Lin), logarithmic (Log), and exponential (Exp) models. In the {exponential} case, all parameter distributions are nearly Gaussian and unimodal, indicating a single, well-defined solution favored by the observed data. For the {logarithmic} model, the posteriors for \( \beta \) and \( Q_0 \) deviate from a perfect Gaussian shape (bimodal), reflecting greater uncertainty and potential degeneracies. The {linear} model shows the most pronounced issues: its \( \alpha \) and \( Q_0 \) distribution is distinctly bimodal, suggesting two separate solutions and its \( \alpha \) posterior is very broad. This bimodality and the associated parameter degeneracies lower the model's Bayesian evidence relative to the others. This visualization highlights why the exponential model is statistically favored, followed by the logarithmic and then the linear model.

\section{Summary and outlook}\label{sec:9}

We employed Bayesian inference with three modified gravity $f(Q)$ models linear, logarithmic and exponential by integrating astrophysical constraints from NICER mass-radius measurements. The NICER data includes PSR J0030+0451, PSR J0740+6620, PSRJ0437+4715, and PSRJ0614+3329. For the first time in the literature on $f(Q)$ gravity, this analysis systematically confronts the theory with neutron star observables in the strong field regime.

The exponential model appeared as the statistically preferred choice, supported by multiple lines of evidence: its \( Q_0 \) parameter is the most tightly constrained ($-0.35^{+0.20}_{-0.36}$), its posterior distributions of all parameters are nearly Gaussian and unimodal.   In contrast, the logarithmic model shows broader parameter uncertainties and mild degeneracies, particularly for \( \beta \) ($1.11^{+2.52}_{-3.02}$). While it predicts the highest maximum neutron-star mass (\( 2.36\,M_\odot \)). The linear model exhibits pronounced bimodality in its \( \alpha \) and \( Q_0 \) posterior and the broadest  \( \alpha \) distribution, reflecting strong parameter degeneracies that reduce its statistical credibility. The tidal deformability $\Lambda_{1.4}$ and radius $R_{1.4}$ for exponential model found to be $156.95^{+84.02}_{-41.73}$ and $11.27^{+0.53}_{-0.36}$, respectively.One of the most significant findings of this work is that all three observationally constrained $f(Q)$ models robustly predict maximum neutron star masses exceeding $2.5\,M_{\odot}$. The corresponding $95\%$ confidence regions of the mass--radius relation consistently enter the lower mass gap ($\sim 2.5$--$5\,M_{\odot}$), a regime where the nature of compact objects remains unresolved. This result is not imposed by construction but emerges directly from the Bayesian-constrained parameter space, demonstrating that the modified gravity effects enhance the high-density gravitational support and allow substantially heavier stable configurations than typically expected within standard GR frameworks. 
The presence of stable neutron stars in the mass-gap region therefore constitutes testable prediction of the present $f(Q)$ models. Future precise mass measurements in this regime could provide a powerful discriminator between GR and extended gravity scenarios.

Model comparisons based on the computed Bayes factors confirm this hierarchy. The logarithmic Bayes factors, \( \ln(\text{BF}) \), decisively favor the exponential model over the logarithmic and linear alternatives, with values of \(-0.35\) and \(-1.28\), respectively. According to the established scale, this indicates strong evidence for the exponential model. Therefore, the exponential \( f(Q) \) model emerges as the best constrained and most statistically supported description of neutron star structure among the three, highlighting its viability as a modified gravity candidate in the strong-field regime.

In the present analysis we adopted a single baseline dense-matter description (DDME2) to reduce nuclear-physics systematics and focus on the impact of the $f(Q)$ sector, with the NICER likelihood and inferred maximum mass explicitly conditioned on this choice. A natural next step is to quantify the robustness of the inferred gravity parameters against EoS uncertainties by performing Bayesian inference for alternative representative nuclear matter EoSs (RMF-based or hybrid), and performing Bayesian model comparison across EoSs in analogy with our Bayes-factor ranking of the $f(Q)$ models. More generally, one can carry out a joint (hierarchical) Bayesian inference in which the $f(Q)$ parameters and a flexible EoS parameterization (e.g., piecewise-polytropie or spectral representations) are sampled simultaneously, thereby marginalizing over the EoS and directly addressing the known degeneracy between strong-field gravity and microphysics; such joint gravity-EoS strategies have proven informative in other modified-gravity contexts. Extending the data evidence beyond NICER (e.g., incorporating experimental nuclear data, empirical nuclear inputs, and theoretical predictions ) would further help break degeneracies and tighten constraints on both sectors.

\textbf{Data availability:} There are no new data associated with this article.\\

\section*{Acknowledgements}S.P. acknowledges the Neutron Star Meeting 2025 held at the Institute of Mathematical Sciences (IMSc), Chennai, India, that motivates this study. NP and KZ acknowledge support by the CUHK-Shenzhen University development fund under grant No. UDF01003041 and UDF03003041, and Shenzhen Peacock fund under No. 2023TC0179, Ministry of Science and Technology of China under Grant No. 2024YFA1611004, NSFC fund under No. 92570117. We also thank Dr. Tuhin Malik, Dr. Sk Md Adil Imam and Asit Karan for technical discussions.

 \bibliographystyle{apsrev4-1}
  \bibliography{ref}

\end{document}